# Prediction of Brain States of Concentration and Relaxation in Real Time with Portable Electroencephalographs

**Zhen Li, Jianjun Xu, Tingshao Zhu**

## Abstract

With the acceleration of our pace of life, our brain states vary greatly every day. We usually find it so hard to adjust our brain states between concentration and relaxation flexibly that we cannot work and rest effectively. Therefore, a convenient and real-time concentration-relaxation feedback system should be established so as to show our current brain state and help us adjust our brain state more flexibly. In our research, we attempt to help people recognize their brain state--concentrating or relaxing more conveniently and in real time with a bio-feedback system. Considering the inconvenience of wearing traditional multiple-electrode electroencephalographs, we choose Muse to collect data which is a portable headband launched lately with a number of useful functions and channels and it is much easier for the public to use. Besides, traditional online analysis did not focus on the synchronism between users and computers and the time delay problem did exist. To solve the problem, by building the Analytic Hierarchy Model, we choose the two gamma wave channels of F7 and F8 as the data source instead of using both beta and alpha channels traditionally. Using the CSP algorithm and the SVM model, the channels we choose have a higher recognition accuracy rate—93.4% and smaller amount of data to be dealt with by the computer than the traditional ones. Furthermore, in order to acquire better synchronism, we make use of the Feedforward Neural Network model to predict users' brain state in 0.5 seconds whose recognition accuracy rate reaches 87.9%. Finally, we design a plane program in Python where a plane can be controlled to go up or down when users concentrate or relax. The SVM model and the Feedforward Neural Network model have both been tested by 12 subjects and they give an evaluation ranging from 1 to 10 points. The former gets 7.58 points while the latter gets 8.83, which proves that the time delay problem is improved once more.

# Introduction

With the acceleration of our pace of life, our brain states vary greatly every day.

When we intend to have a rest after work or study, it is difficult for us to make our brain calm down and relax immediately. On the other hand, when we work or study, we sometimes cannot concentrate so well that the work efficiency may be rather low. In a word, we usually find it hard to adjust our brain state flexibly in order to suit different conditions. Therefore, an effective and real-time concentration-relaxation feedback system should be established so as to show our current brain state, which can help us adjust our brain state between concentration and relaxation flexibly. In the previous researches, researchers had attempted to help people concentrate and relax using kinds of electroencephalographs. However, in general, two problems existed which we will solve in our research. The first problem is that the electroencephalographs they used always had many electrodes and they were complex for users to wear. In our research, we will make use of Muse--a portable headband launched by InteraXon Company in 2014 to collect data of brain waves. Muse provides useful functions and channels which are more beneficial for us to build a brain-state-recognition model with CSP and SVM algorithms. Besides, another problem is that the previous researches always focused on offline analysis to recognize the brain states. Despite a few researches relating to online analysis, the synchronism between users and computers was usually poor and the time delay problem existed, which meant their feedback systems could not show users' brain state in time. To solve the time delay problem, firstly, in the premise of a high recognition accuracy rate, we intend to decrease the amount of data to be dealt with by the computer. With the help of the Analytic Hierarchy Model, we choose the two gamma wave channels of F7 and F8 as the data source instead of using both beta and alpha channels traditionally and they have a higher recognition accuracy rate—93.4% and smaller amount of data to be dealt with by the computer. Moreover, we make use of the Feedforward Neural Network model to replace the SVM model to predict users' brain states in 0.5 seconds, helping improve the time delay problem. Finally, we design a plane program in Python where a plane can be controlled to go up or down when users concentrate or relax. The SVM model and the Feedforward Neural Network model have both been tested by 12 subjects with satisfactory evaluation and the latter is generally thought to have better synchronism.

# Related work

As far as we know, a number of researchers have focused on recognition of brain states of concentration and relaxation in recent years. Xueni Li and Yufeng Wang et al.[2] studied EEG biofeedback on treatment of attention deficit hyperactivity disorder. They trained subjects to concentrate with the help of feedback of various oscillograms of EEG in real time. Yanbing Zhang[3] researched on the feature of EEG, myoelectricity and mood when subjects relaxed and gave the synthetic index of the brain state after several minutes' data collection. Huiche Science Company invented the mind control mini car[5] in 2014 which aimed to help people concentrate better. Despite the marvelous contribution they made, their methods had some limitations to some extent. Firstly, the electroencephalographs they used always had many electrode poles which were complex for users to wear. In our research, we make use of Muse--a portable headband with only four electrodes to collect data of brain waves and it should be more likely to be used by public. Furthermore, in most previous researches, the bio-feedback systems could not show users' brain state in time and the synchronism between users and computers was usually poor. However, by choosing proper channels and building a new model, we improve the time delay problem so that users can know their brain state in real time and more clearly.

# Methods

## 1. The portable electroencephalograph—Muse

In our research, considering the convenience to wear, varieties of data channels and the useful functions, we made use of the portable electroencephalograph—Muse [8] for data collection and tests which was proved to be successful in the following experiment.

Muse was first launched in 2014 by InteraXon Company. Traditionally, electroencephalographs with over 10 electrode poles are usually used in the experiment of dealing with brain signals. These electroencephalographs are generally difficult for subjects to wear and cannot be widely used in people's daily life. Furthermore, traditional electroencephalographs can merely provide limited data source such as EEG, alpha and beta channels. On the other hand, Muse has only 4 electrode poles to measure brain waves of four parts—F7, F8 and the back of ears as reference. The places of the corresponding electrode poles are shown in Figure 2. It is much easier for people to wear instead of using electrode cream before and transmits data collected to the computer via Bluetooth wirelessly. Apart from its convenience to wear, Muse also provides multiple data channels including not only the channels listed above but also gamma wave channels, which will be proved to be more advantageous than others in our experiment. Moreover, Muse provides functions to filter the impure brain waves including the power frequency automatically, which helps increase the signal-noise ratio. Therefore, we use Muse to collect the data of brain waves in the experiment considering its convenience to wear, varieties of data channels and the useful functions.

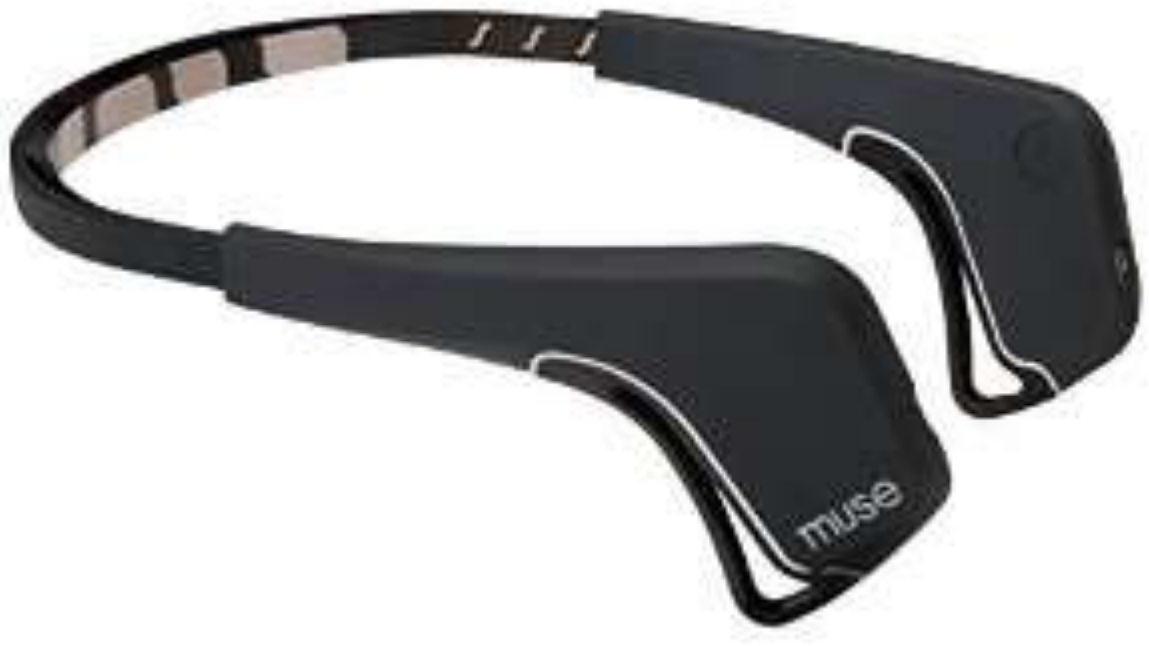

Figure 1 Muse

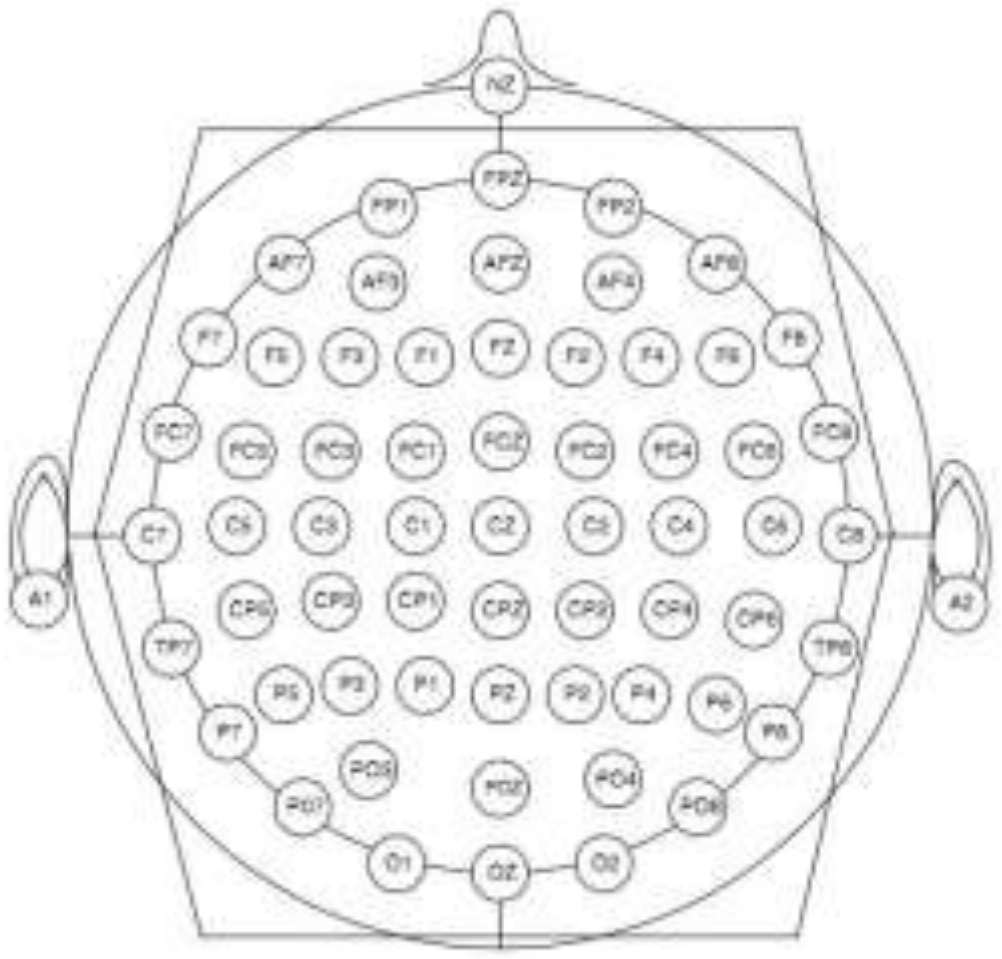

Figure 2 Location of electrode poles

## 2. Support Vector Machine

In order to choose the proper channels as the data source, we needed to consider the recognition accuracy rate and the number of the channels selected. Firstly, we needed to build a proper classifier to get recognition accuracy rates of different channels for comparison. Like many researchers who study recognition of brain states, we also made use of the Support Vector Machine model[9] to build our classifier.

Due to the complexity and variability of brain waves, the traditional classifiers, which are mainly based on a large sample, lack the stability and accuracy to predict the change of brain waves. As a result, we applied the Support Vector Machine to project the eigenvectors we had extracted to a higher dimensional space, where it is simpler to discriminate the classes. Four kernel functions are mainly used in SVM, and we chose radial basis function for our experiment.

## 3. Feature extraction

After we regarded SVM as the algorithm to build the classifier, the following task for us was to extract the feature of data of different brain states to increase the recognition accuracy rate. According to some related documents, the CSP algorithm[11] can extract the feature of different brain waves effectively. Therefore, we made use of the CSP algorithm to extract the feature of data of different brain states. In the following part, we will first make a short introduction about the CSP algorithm. After we got a series of recognition accuracy rates of different channels using CSP and SVM algorithms, we used the diagram analysis and the Hierarchy Analysis model[12] to choose the most suitable channels.

## 3.1 CSP algorithm

Since brain signals are reflections of numerous active brain sources, the common sources do exist when we attempt to concentrate and relax. Therefore, the discrepancy between the two states is bound to be influenced and the work of extracting the feature is shown to be complicated. In order to separate the related sources from the common sources, we use the Common Space Pattern to calculate spatial filters for detecting ERD/ERS effects. Subsequently, we define the eigenvector and acquire a new data set to train the Support Vector Machine.

### The concrete process of dealing with data with CSP in the experiment

1. We obtained a number of groups of data of concentration and relaxation from different subjects.

2. We figured out the spatial filters $W_T$ and $W_R$ for each group. Then the average spatial filters $W_T^A$ and $W_R^A$ could be got.

3. As for a data set $X^T$ with $K$ rows and $N$ columns, we let

$$H_T = W_T^A \times X \quad (3.1)$$

$$H_R = W_R^A \times X \quad (3.2)$$

where $H_T \in R^{1\times K}$ and $H_R \in R^{1\times K}$.

4. We combined $H_T$ and $H_R$ into a $2K$ -dimension eigenvector and it was regarded as the input of SVM.

## 3.2 The choice of channels

As far as we are concerned, the types of data that traditional electroencephalographs provide are always limited. They can only provide data of EEG, alpha waves, beta waves and some other basic data types. On the other hand, Muse can provide us with various types of data including EEG, gamma, beta, alpha, acc, FFT and etc, which contain more information to recognize the brain states. Researchers who have studied the classification of the brain waves between concentration and relaxation states always use the data from EEG, alpha wave or beta wave channels. In our paper, we built the Analytic Hierarchy Model to choose the optimum channels so as to increase the classification accuracy rate and reduce the complexity of data. F7 and F8 electrode poles were the poles we chose and we intended to choose channels among gamma, beta and alpha wave ones. Thus, the number of the channels we chose should be 2, 4 or 6. As the increase of channels might increase time to run the computer program, we could not choose too many channels to build the classifier. Our purpose was to predict the brain state both accurately and in real time. Therefore, we took the accuracy rate of the prediction, the number of channels and the preknowledge of the types of data into consideration and finally chose the two gamma channels of F7 and F8 as the data source.

### 3.2.1 Diagram Analysis

We chose 100 items of data of different brain waves from a subject both in the concentration state and relaxation state and then we made 3 scatter diagrams where circles stood for concentration states and triangles for relaxation ones.

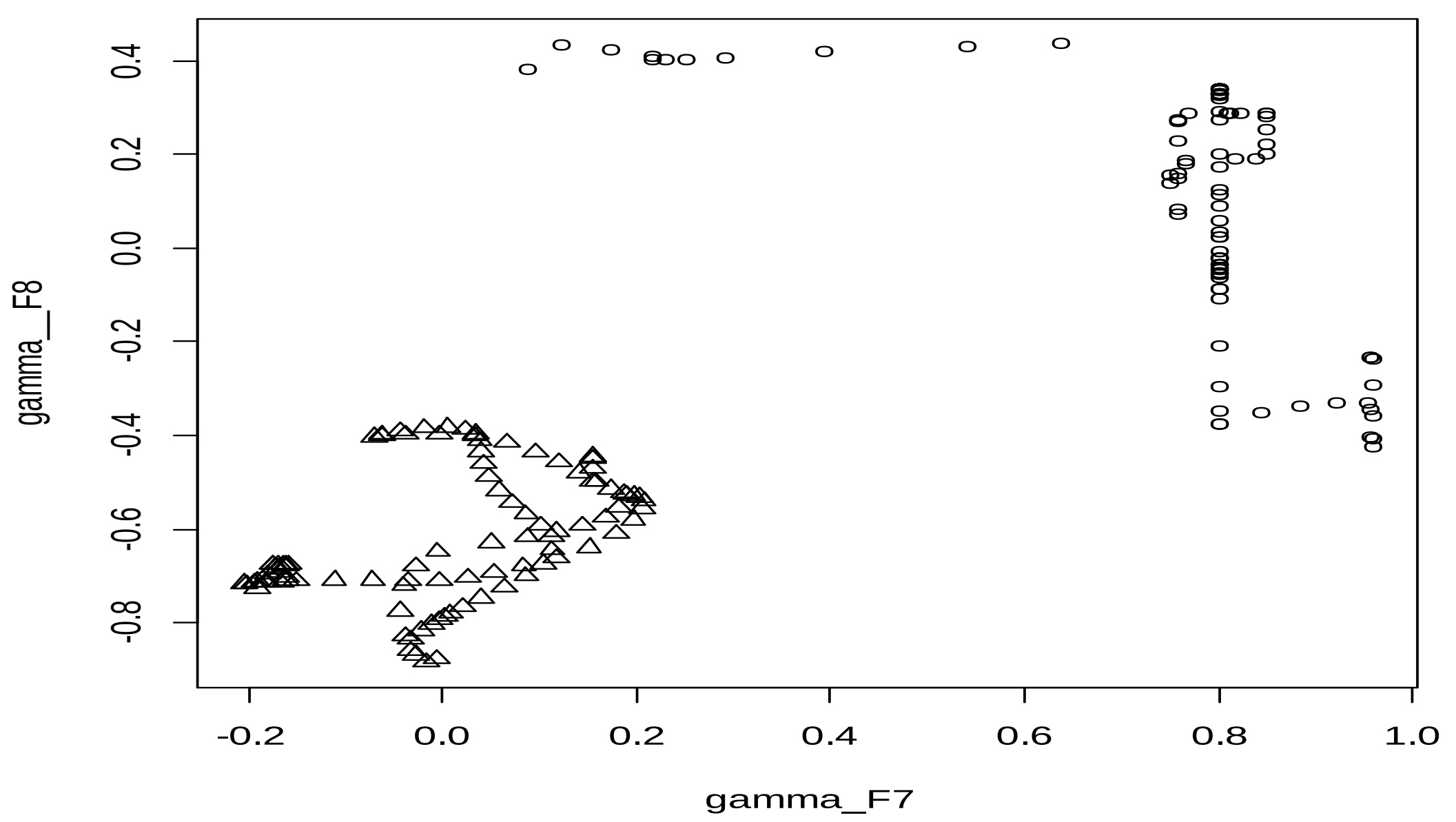

Figure 3 Gamma power of concentration and relaxation state from F7 and F8

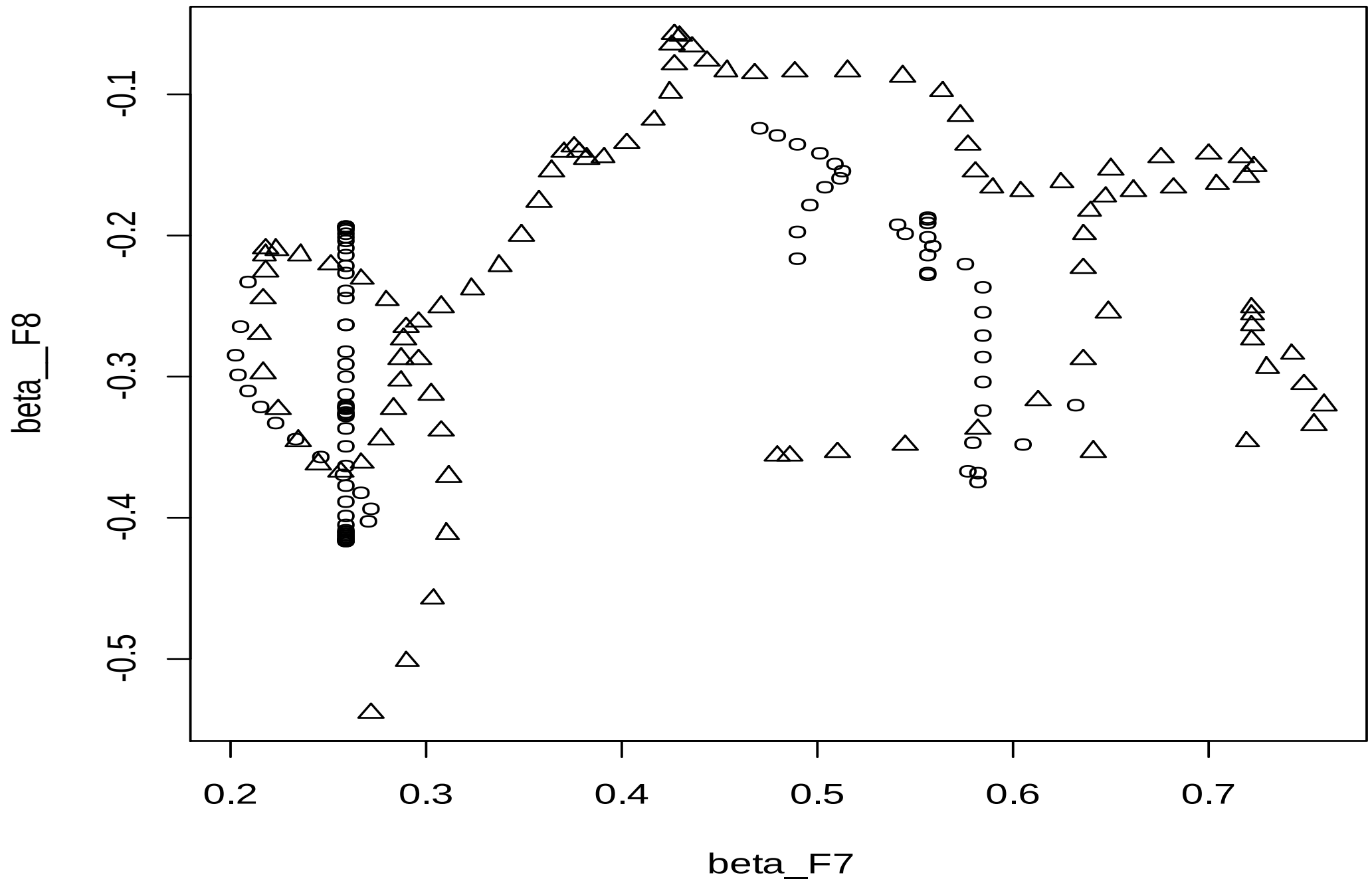

Figure 4 Beta power of concentration and relaxation state from F7 and F8

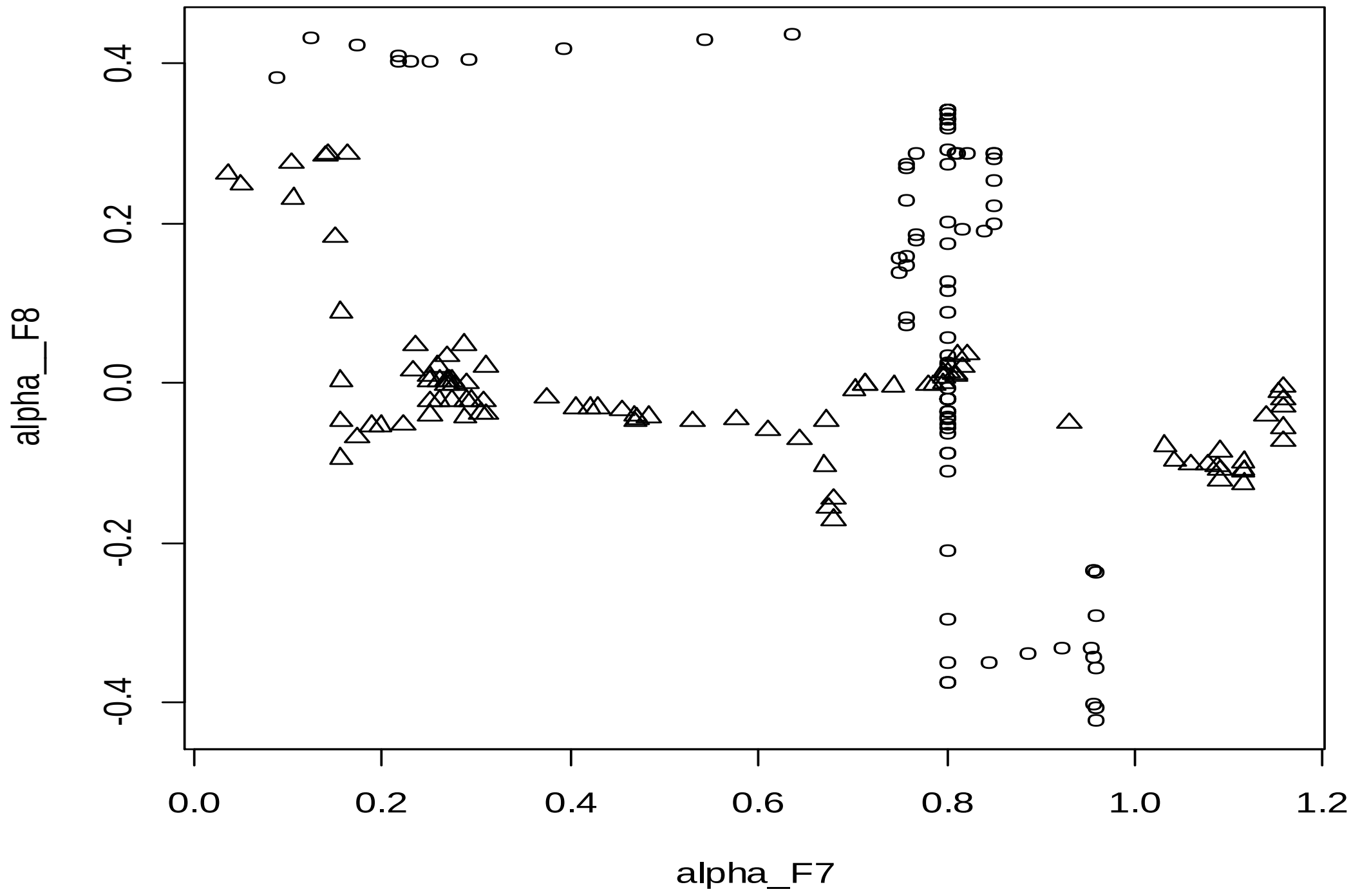

Figure 5 Alpha power of concentration and relaxation state from F7 and F8

From the three scatter diagrams above, we found the data of two gamma wave channels could be discriminated most evidently and we had a basic idea of the choice of channels.

### 3.2.2 Analytic Hierarchy Model

We knew from some psychological papers[6][7] that when a person concentrates, beta and gamma waves detected from F7 and F8 will increase and they will decrease when the person relaxes. Hence, the data of beta and gamma detected from the forehead has a good discrimination property between the concentration and the relaxation state. The data of alpha waves detected from the back of the head has also a good discrimination property but the electroencephalograph we used had only detectors attached to the forehead and the back of the ears. Therefore, we supposed $c_2$ equaled to 0.5 when alpha channels were used otherwise it equaled to 1. Furthermore, the smaller the number of channels is, the less time it will take to deal with the data and run the computer program so that the classifier is more likely to keep in real time. This is what we expect. Thus, we regarded the reciprocal of the number of channels as a factor for us to choose certain channels.

Then, we divided the problem into two hierarchies-the target hierarchy and the factor hierarchy and built the Analytic Hierarchy Model.

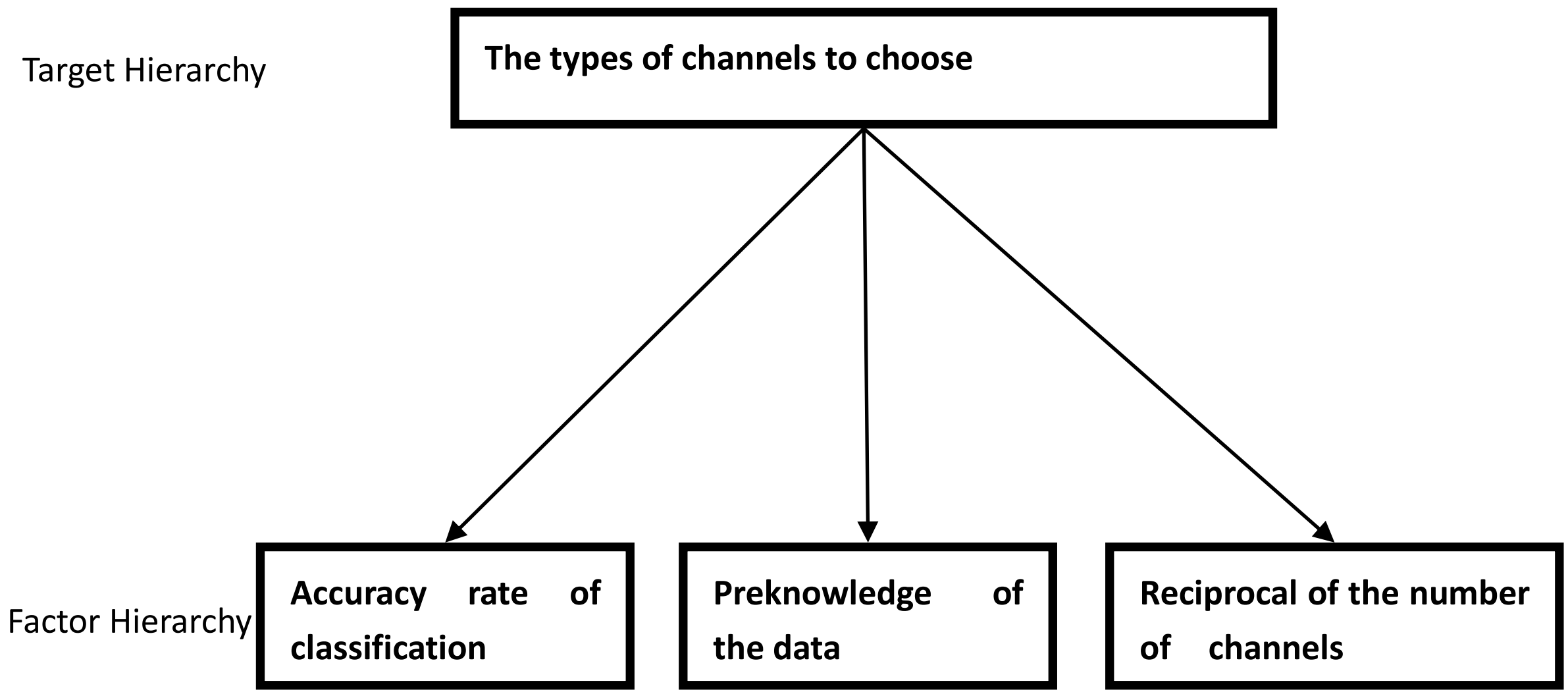

Figure 6 Basic framework of the Analytic Hierachy model

Since the accuracy rate of classification weighed most in the evaluation with no doubt, we gave this factor a rather big weight. Moreover, the number of time to run the computer program differed a little when the number of channels varied; so the factor did not weigh much. Finally, we were not sure that the preknowledge obtained from the Internet suited to our experiment well and we should rely much more on the result of the experiment than the preknowledge; thus, the preknowledge of data only had a slight weight. Therefore, we defined $a_{ij}$ to show the relative importance between $i$ and $j$ as following:

$$\begin{aligned} & a_{12} = 8 \\ & a_{13} = 3 \\ & a_{23} = 3 \\ & a_{ii} = 1 \\ & a_{ij} = \frac{1}{a_{ji}}, i, j = 1, 2, 3 \end{aligned} \qquad (3.3)$$

Then we gave the corresponding comparison matrix

$$A = \begin{bmatrix} 1 & 8 & 3 \\ \frac{1}{8} & 1 & \frac{1}{3} \\ \frac{1}{3} & 3 & 1 \end{bmatrix} \qquad (3.4)$$

With standardization, the eigenvector of the maximal eigenvalue of $A$ was $(0.682, 0.082, 0.236)^T$

Hence, we got the coefficient of the evaluation and let

$$Q = 0.682c_1 + 0.082c_2 + 0.236c_3 \qquad (3.5)$$

As the scale of the three factors differed, they should be standardized through being divided by their own maximum number in their range. We could easily know the maximum of the three factors were 1,1,0.5 seperately. From the data we had collected from 10 subjects, we figured out Q for each choice of channels and got the mean values as the following table shows:

| | **Gamma** | **Beta** | **Alpha** | **gamma and beta** |
|---|---|---|---|---|
| **Q** | 0.948 | 0.892 | 0.764 | 0.885 |

Table 1 The value of Q for different channels

As a matter of fact, we could not quantify the importance of the three factors specifically but we had to estimate the relative importance $a_{ij}$ among them initially otherwise the model would not work. Fortunately, the result of the experiment revealed that no matter how the relative importance varied, the channels we should choose were invariable.

Therefore, we chose the two gamma channels of F7 and F8 as the data source which was proved to be successful in the experiment later.

# 4. The plane program

After the proper channels were chosen, we needed to devise suitable software to show people's current brain state of concentration or relaxation in real time. A plane program where a plane can be controlled to go up or down when users concentrate or relax was our final choice.

The purpose of the plane program was to show subjects' brain states of concentration or relaxation in real time. Above all, we made use of C++ to write a program to transmit data of brain waves collected by Muse to Python in real time. Subsequently, we transferred the format of the classifiers we made in R Language to the one that was compatible to Python. The output frequency for the gamma wave was 10hz which meant we could obtain 10 rows of data from F7 and F8 per second. As for the SVM model, we regarded 5 rows of data as a whole and get the eigenvector by CSP algorithm as the input of the corresponding classifier. As for the Feedforward Neural Network model, we combined 5 rows of data to a 10-dimension vector as the input of the corresponding classifier. The input was generated every 0.1 second which consisted of data of a subject's current state and 0.5 seconds forward. The plane program was written in Python. If the input was labeled as 1 by the classifier, the plane would rise up a bit and if the input was labeled as -1, it would go down a bit. According to the above, the plane would execute a command every 0.1 second to show subjects' brain states of concentration and relaxation in real time.

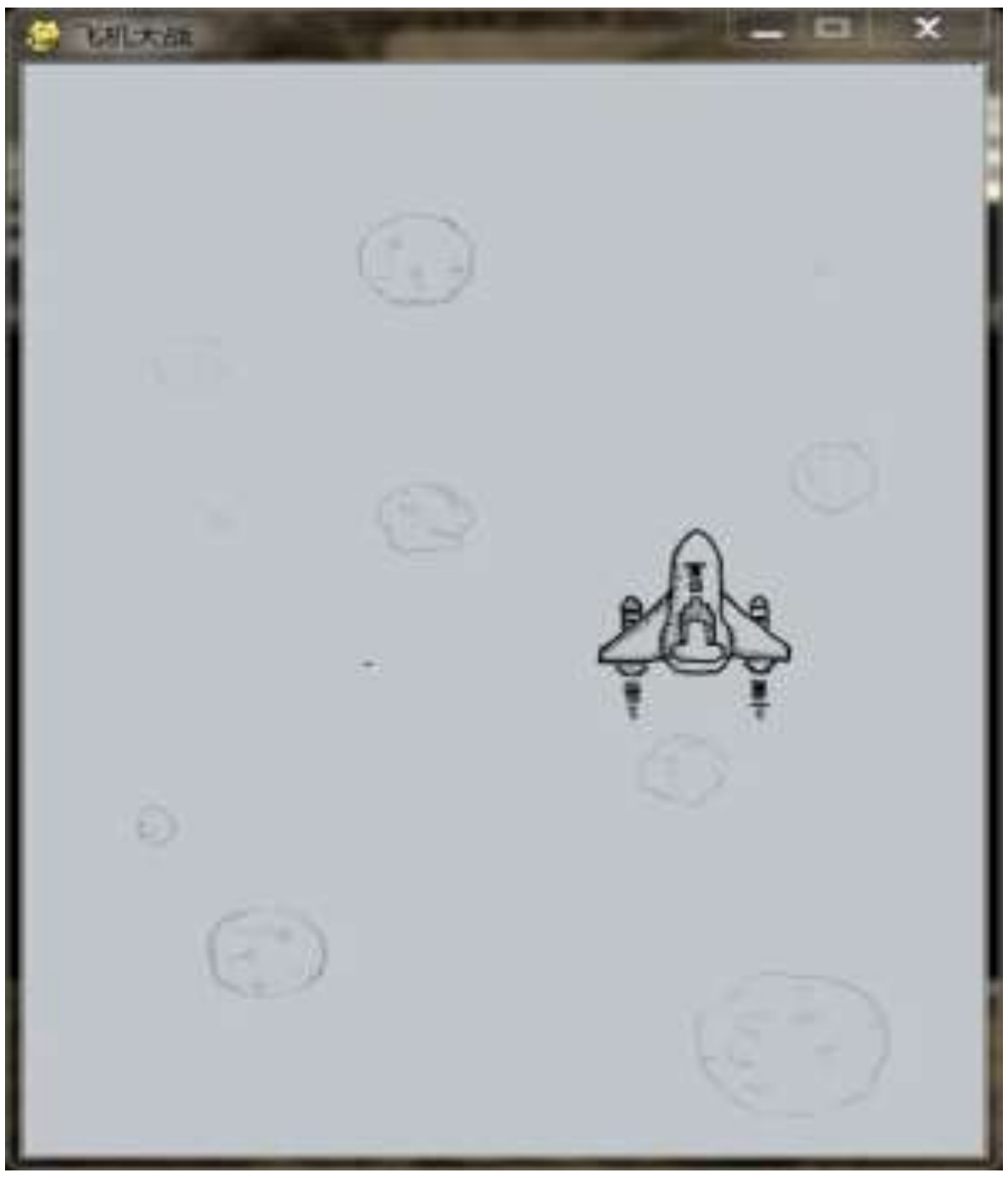

Figure 7 The plane program interface

# 5. Experiment I

After the channels were chosen and the plane software was made, we needed to check out if the feedback system could recognize people's brain state in real time. A SVM classifier should be made using training samples first. Then the users' data of brain waves would be labeled 1 or -1 by the classifier. The label would be the instruction which controlled the plane to move. When the subject concentrated, the plane would rise up and when he relaxed, it would go down.

## 5.1 Experiment purpose

We needed to collect the data of brain waves when subjects concentrated and relaxed in order to make a general SVM classifier to suit every user. With the help of the classifier, the software mentioned above could predict the subject's brain state after the head band was connected to the computer.

## 5.2 Experiment subjects

In the experiment, we collected data of brain waves from 20 adults ranging in age from 19 to 30. Half of the subjects were male and all of the subjects came from China. 65% of the subjects had completed at least some college and 35% of them were undergraduates.

Besides, we tested 8 subjects for the plane program and 6 of them were from the 20 mentioned above. Three quarters of the subjects were male and all of the subjects were from China. 25% of the subjects had completed at least some college and the rest of them were undergraduates.

## 5.3 Experiment content

As the power frequency interfered with the brain waves greatly, we had to filter the noise with the help of certain functions in Muse. This was the pretreatment of the experiment. Before the experiment, we divided the subjects into two groups: A and B. At the beginning of the experiment, a subject's forehead should be cleaned with an alcohol cotton ball. Then we helped him wear the headband and made it fit for him. After that, we asked subjects in Group A to see a photo of a plane on the screen and imagine it was moving around quickly (some other complicated movements were also permitted). The process would be done 20 times and 10 seconds for each. Then the subjects might have 2 minutes' rest. After the rest, the subjects were asked to close their eyes and relax themselves. The process would be done 20 times likewise but 15 seconds for each. The data of the subjects' brain waves were recorded and saved in csv files. On the other hand, the subjects in Group B were asked to finish the same task by the opposite order. After each experiment, we would make a classifier using the data and figured out the accuracy rate. We would choose 30 of the 40 data sets randomly as the training samples while the rest were the test samples. If the accuracy rate was over 80 percent, we assumed that the result of the test could be accepted. Otherwise, we would analyze the reasons and attempted to improve the experiment procedure.

## 5.4 Time domain Analysis

As we have mentioned before, we chose the two gamma wave channels that were attached to the forehead as our data source. The output frequency for the gamma wave was 10hz which meant we could obtain 10 rows of data from the two channels per second. However, despite the filter for the power frequency, not all of the data saved could be effective. According to some related documents[17], we would make use of 5-second data in a 10-second test. We regarded data in different time periods as effective supposedly, made separate classifiers and figured out the mean accuracy rate of each classifier as the following table shows.

| | 3-7s relaxation | 4-8s relaxation |
|---|---|---|
| 1-5s concentration | 86.5% | 91.5% |
| 2-6s concentration | 88.1% | 93.4% |

Table 2 The mean accuracy rates of different time periods

As a consequence, we regarded the 2-6s data of concentration and the 4-8s data of relaxation as effective.

# 6. Feedforward neural network to solve the time delay problem

As for the SVM model, although the majority of the subjects could control the plane by concentrating and relaxing, they told us that the plane would always respond to their thought 1 or 2 seconds later. In order to solve the time delay problem, we applied the Feedforward Neural Network model[13] to predict a subject's brain state in 0.5 seconds. After the construction of the model, we asked 12 people to control the plane by both the SVM and the Feedforward Neural Network model. Most of them admitted they could control the plane more flexibly by the Neural Network model and the time delay problem had been improved to some degree.

In our Feedforward Neural Network Model, the input layer included the data of a subject's current state and 0.5 seconds forward. Since we chose the two gamma wave channels as our data source and the output frequency for the gamma wave was 10hz, we could get 10 values in the 0.5-second-period. Thus, we combined these values to a 10-dimension vector as the input. The output layer included the brain state in 0.5 seconds. If the brain state was recognized as concentration, the output was 1. If the brain state was recognized as relaxation, the output was -1. In our experiment, we set learning rate as 0.01, hidden layer as “transient”, output layer as “purelin” and 10 hidden-layer nodes.

# 7. Experiment II

After the construction of the Feedforward Neural Network model, we asked 12 people to control the plane by both the SVM and the Feedforward Neural Network model and they finally gave an evaluation ranging from 1 to 10 points. The former model got 7.58 points while the latter got 8.83, which proved that the time delay problem was improved to a certain extent.

## 7.1 Experiment purpose

As some subjects who had finished the first experiment told us that the plane always responded to their thought 1 or 2 seconds later, we intended to make use of the Feedforward Neural Network model to improve the problem. We hoped to show that the Feedforward Neural Network model could help users control the plane more flexibly than the SVM model and it could solve the time delay problem to some degrees.

## 7.2 Experiment subjects

In the experiment, we collected data of brain waves from 12 adults and tested them with both the SVM and the Feedforward Neural Network model. The subjects’ ages ranged from 19 to 25. 33% of the subjects were male and all of the subjects came from China. Two thirds of the subjects had completed at least some college and the rest of them were undergraduates.

## 7.3 Experiment content

As the power frequency interfered with the brain waves greatly, we had to filter the noise with the help of certain functions in Muse before the experiment. At the beginning of the experiment, we helped the subject choose a suitable SVM classifier from the 5 mentioned in Experiment I and told him to control the plane flexibly by concentrating and relaxing. Subsequently, we asked him to raise the plane for 5-10 seconds and then drove the plane down for 5-10 seconds and told him to repeat this task for 2 minutes totally. After saving the 120s data in a txt file, we used the chosen SVM classifier to label the data. We regarded the 120s data as the input of the neural network and the label as the output. After the classifier based on the Feedforward Neural Network model had been built, we told the subject to control the plane again.

# Results

## 1. The recognition accuracy rate and mind control effect of the SVM model

In experiment I, we tested 20 subjects and only 1 data set of them did not qualify our standard. Four order cross validation was used to obtain the recognition accuracy rates so that the results could be regarded as reliable. The overall recognition accuracy rates are shown in Table 3 and the mean accuracy rate was 93.4% which proved that the data we got was relatively pure and effective. As different people's brain waves ranged differently, we chose certain data sets of the 20 and made five classifiers totally to suit each user. After that, we found 8 subjects and told them to control the plane program written in Python by concentrating and relaxing. After being trained and adjusting, 7 of them could control the plane to go up or down flexibly which proved that the SVM model could predict whether a subject was concentrating or relaxing well.

| Subject | Concentration | Relaxation | All |
|---|---|---|---|
| 1 | 100% | 100% | 100% |
| 2 | 84.2% | 95.9% | 90% |
| 3 | 80.4% | 77.5% | 79% |
| 4 | 100% | 100% | 100% |
| 5 | 94.3% | 90.5% | 92.4% |
| 6 | 89.7% | 98.1% | 93.9% |
| 7 | 98.7% | 84.5% | 91.6% |
| 8 | 94.7% | 88.2% | 91.5% |
| 9 | 91.9% | 96.7% | 94.3% |
| 10 | 100% | 100% | 100% |
| 11 | 96.4% | 97.2% | 96.8% |
| 12 | 99.8% | 95.2% | 97.5% |
| 13 | 96% | 99.2% | 97.6% |
| 14 | 93.4% | 96.2% | 94.8% |
| 15 | 99.6% | 98.4% | 99% |
| 16 | 82.3% | 87.7% | 85% |
| 17 | 86.4% | 83.6% | 85% |
| 18 | 93.9% | 91.1% | 92.5% |
| 19 | 90.3% | 83.9% | 87.1% |
| 20 | 100% | 100% | 100% |
| Average | 93.6% | 93.2% | 93.4% |

Table 3 The recognition accuracy rates of the SVM model

## 2. The recognition accuracy rate and mind control effect of the Feedforward Neural Network model

In experiment II, we collected data of 12 subjects and tested them with the Feedforward Neural Network model. Four order cross validation was used to obtain the recognition accuracy rates so that the results could be regarded as reliable. The overall recognition accuracy rates of the state in 0.5s are shown in Table 4 and the mean accuracy rate was 87.9% which was a bit lower than that of the SVM model. We asked the 12 subjects to control the plane program and we found that the feedback system had rather good synchronism although the recognition accuracy rate was not as high as that of the SVM model. The plane's sensitivity was rather high and it was more likely to be influenced by the change of brain waves of the subjects.

| Subject | State in 0.5s |
|---|---|
| 1 | 90.2% |
| 2 | 92.1% |
| 3 | 83.6% |
| 4 | 82.4% |
| 5 | 85.8% |
| 6 | 84.3% |
| 7 | 89.9% |
| 8 | 92.8% |
| 9 | 88.5% |
| 10 | 90.5% |
| 11 | 87.8% |
| 12 | 86.9% |
| Average | 87.9% |

Table 4 The recognition accuracy rates of the Feedforward Neural Network model

## 3. The comparison between the SVM and the Feedforward Neural Network model

We tested 12 subjects and asked them to evaluate the two models. Subjects could offer an evaluation ranging from 1 to 10 points for the two models where 10 points meant they could control the plane very flexibly and 1 point meant they could not control it at all. The overall points of the 12 subjects' evaluations were as following:

| Subject | SVM model | Neural Network model |
|---|---|---|
| 1 | 9 | 10 |
| 2 | 9 | 10 |
| 3 | 7 | 7 |
| 4 | 9 | 10 |
| 5 | 7 | 8 |
| 6 | 5 | 7 |
| 7 | 8 | 10 |
| 8 | 7 | 8 |
| 9 | 8 | 9 |
| 10 | 7 | 9 |
| 11 | 9 | 9 |
| 12 | 6 | 8 |
| Average | 7.58 | 8.83 |

Table 5 The evaluations of the SVM and the Feedforward Neural Network model

From the table above, we found the average points of the Feedward Neural Network model was higher than those of the SVM model although the Neural Network model had a lower recognition accuracy rate—87.9% than 93.4%. Consequently, we could make a conclusion that the Feedforward Neural Network model improved the synchronism between the subject and the computer program indeed and the time delay problem had been improved to some degrees.

## Discussion

From the results above, users can know about their current brain state of concentration or relaxation conveniently with Muse and promptly with the help of the Feedforward Neural Network model. Like previous researches, our feedback system can recognize subjects' brain state of concentration or relaxation online. However, compared to a number of researches, the electroencephalograph with fewer electrodes we use is more portable and more likely to be used widely. Furthermore, although some feedback systems such as the mind control mini car can recognize the brain state of concentration or relaxation online, the synchronism of them is always limited and users cannot know their brain state in time. With the help of the Feedforward Neural Network model, we improve the time delay problem to some degrees and our feedback system has better synchronism. In our research, we make use of four order cross validation to get the recognition accuracy rates so that the results above can be regarded as believable. Honestly, our feedback system has some limitations. The main limitation is that we do not define concentration and relaxation clearly which means that there are not specific critical values to discriminate the brain waves of concentration and relaxation. Instead, our feedback system can only give a relative recognition of one's brain state. For instance, when a subject concentrates, our system may recognize his brain state as concentration one. However, in a case, if he is not as concentrative as he was just now, the feedback system may recognize his brain state as relaxation one even though it does not match to the standard of relaxation states that other researchers have defined. Nevertheless, our purpose is to help people adjust their brain state between concentration and relaxation flexibly. Although our system can only give a relative recognition, it is useful enough for users to change their brain state from a concentrative one to a less concentrative one and finally to a relaxing one. The opposite effect can be obtained as well. As a consequence, the limitation does not have an obvious influence on our aim and the results of the experiment.

# Conclusions

Our research aims at helping people recognize their brain state--concentrating or relaxing more conveniently and in real time. Considering the inconvenience of wearing traditional multiple-electrodes electroencephalographs, we choose Muse to collect data which is a portable headband launched lately with a number of useful functions and channels and it is much easier for the public to use. Besides, traditional online analysis did not focus on the synchronism between users and computers and the time delay problem did exist. To solve the problem, we choose the two gamma wave channels of F7 and F8 as the data source instead of using both beta and alpha channels traditionally, which has a higher recognition accuracy rate and smaller amount of data to be dealt with by the computer. Furthermore, we make use of the Feedforward Neural Network model to predict users' brain state in 0.5 seconds, helping improve the time delay problem. The traditional model and the new model have both been tested by 12 subjects and the latter is generally thought to have better synchronism. In conclusion, users can recognize their brain state of concentration and relaxation more conveniently and in real time.